\documentclass[10pt,apjl,tighten,iop]{emulateapj}

\usepackage{amsmath}   % for \being{align}...
\usepackage{graphicx}
\usepackage{tabulary}
\usepackage{epsfig}
\usepackage{amssymb}
\usepackage{multirow}
\usepackage{appendix}
\usepackage{natbib}
\usepackage{lineno}
\usepackage{wasysym}
\usepackage[pdftitle={Water ice lines and the formation of giant moons around super-Jovian planets}, colorlinks = true, breaklinks = true,
  citecolor = blue, linkcolor = blue, urlcolor = blue, pdfauthor =
  {Ren\'{e} Heller}]{hyperref}
\newcommand{\beq}[1]{\begin{equation}\label{#1}}
\newcommand{\eeq}{\end{equation}}

\shorttitle{Water ice lines and the formation of giant moons around super-Jovian planets}

\shortauthors{Ren\'e Heller {{\&}} Ralph Pudritz}

\begin{document}

%FRONT MATTER
\title{Water ice lines and the formation of giant moons around super-Jovian planets}

\author{Ren\'e Heller\altaffilmark{1,2} and Ralph Pudritz\altaffilmark{1}}
\affil{Origins Institute, McMaster University, Hamilton, ON L8S 4M1, Canada \\\href{mailto:rheller@physics.mcmaster.ca}{rheller@physics.mcmaster.ca}, \href{mailto:pudritz@physics.mcmaster.ca}{pudritz@physics.mcmaster.ca} }

\altaffiltext{1}{Department of Physics and Astronomy, McMaster University}
\altaffiltext{2}{Postdoctoral fellow of the Canadian Astrobiology Training Program}

%ABSTRACT
\begin{abstract}
Most of the exoplanets with known masses at Earth-like distances to Sun-like stars are heavier than Jupiter, which raises the question of whether such planets are accompanied by detectable, possibly habitable moons. Here we simulate the accretion disks around super-Jovian planets and find that giant moons with masses similar to Mars can form. Our results suggest that the Galilean moons formed during the final stages of accretion onto Jupiter, when the circumjovian disk was sufficiently cool. In contrast to other studies, with our assumptions, we show that Jupiter was still feeding from the circumsolar disk and that its principal moons cannot have formed after the complete photoevaporation of the circumsolar nebula. To counteract the steady loss of moons into the planet due to type I migration, we propose that the water ice line around Jupiter and super-Jovian exoplanets acted as a migration trap for moons. Heat transitions, however, cross the disk during the gap opening within $\approx10^{4}$\,yr, which makes them inefficient as moon traps and indicates a fundamental difference between planet and moon formation. We find that icy moons larger than the smallest known exoplanet can form at about 15 - 30 Jupiter radii around super-Jovian planets. Their size implies detectability by the \textit{Kepler} and \textit{PLATO} space telescopes as well as by the \textit{European Extremely Large Telescope}. Observations of such giant exomoons would be a novel gateway to understanding planet formation, as moons carry information about the accretion history of their planets.
\end{abstract}

%KEY WORDS
\keywords{accretion disks -- planets and satellites: formation -- planets and satellites: gaseous planets -- planets and satellites: physical evolution -- planetÐdisk interactions}

\section{CONTEXT}
\label{sec:context}

While thousands of planets and planet candidates have been found outside the solar system, some of which are as small as the Earth's moon \citep{2013Natur.494..452B}, no moon around an exoplanet has yet been observed. But if they transit their host stars, large exomoons could be detectable in the data from the \textit{Kepler} space telescope or from the upcoming \textit{PLATO} mission \citep{2012ApJ...750..115K,2014ApJ...787...14H}. Alternatively, if a large moon transits a self-luminous giant planet, the moon's planetary transit might be detectable photometrically or even spectroscopically, for example with the \textit{European Extremely Large Telescope} \citep{2014ApJ...796L...1H}. It is therefore timely to consider models for exomoon formation.

Large moons can form in the dusty gas disks around young, accreting gas giant planets. Several models of moon formation posit that proto-satellites can be rapidly lost into the planet by type I migration \citep{1974Icar...21..248P,2002AJ....124.3404C,2006Natur.441..834C,2010ApJ...714.1052S}. The water (H$_2$O) condensation ice line can act as a planet migration trap that halts rapid type I migration in circumstellar disks \citep{2007ApJ...664L..55K,2011MNRAS.417.1236H,2012ApJ...760..117H}, butÊthis trap mechanism has not been considered in theories of moon formation so far. The position of the H$_2$O ice line has sometimes been modeled ad hoc \citep{2010ApJ...714.1052S} to fit the H$_2$O distribution in the Galilean moon system \citep{2003Icar..163..198M}.

An alternative explanation for the formation of the Galilean satellites suggests that the growingÊIo, Europa, and Ganymede migrated within an optically thick accretion disk the size of about the contemporary orbit of Callisto and accreted material well outside their instantaneous feeding zones \citep{2003Icar..163..198M,2003Icar..163..232M}. In this picture, Callisto supposedly formed in an extended optically thin disk after Jupiter opened up a gap in the circumsolar disk. Callisto's material was initially spread out over as much as $150$ Jupiter radii ($R_{\rm Jup}$), then aggregated on a $10^6$\,yr timescale, and migrated to Callisto's current orbital location. Yet another possible formation scenario suggests that proto-satellites drifted outwards as they were fed from a spreading circumplanetary ring in a mostly gas-free environmentÊ\citep{2012Sci...338.1196C}.

We here focus on the ``gas-starved'' model of an actively supplied circumplanetary disk (CPD) \citep{1999SoSyR..33..456M,2002AJ....124.3404C} and determine the time-dependent radial position of the H$_2$O ice line. There are several reasons why water ice lines could play a fundamental role in the formation of giant moons. The total mass of a giant planet's moon system is sensitive to the location of the H$_2$O ice line in the CPD, where the surface density of solids ($\Sigma_{\rm s}$) increases by about factor of three \citep{1981PThPS..70...35H}, because the mass of the fastest growing object is proportional to $\Sigma_{\rm s}^{3/2}$. This suggests that the most massive moons form at or beyond the ice line. In this regard, it is interesting that the two lightest Galilean satellites, Io (at $6.1\,R_{\rm Jup}$ from the planetary core) and Europa (at $9.7\,R_{\rm Jup}$), are mostly rocky with bulk densities $>3\,{\rm g\,cm}^{-3}$, while the massive moons Ganymede (at $15.5\,R_{\rm Jup}$) and Callisto (at $27.2\,R_{\rm Jup}$) have densities below $2\,{\rm g\,cm}^{-3}$ and consist by about 50\,\% of H$_2$O \citep{1999Sci...286...77S}.\footnote{Amalthea, although being very close to Jupiter (at $2.5\,R_{\rm Jup}$), has a very low density of about $0.86\,{\rm g\,cm}^{-3}$ \citep{2005Sci...308.1291A}, which seems to be at odds with the compositional gradient in the Galilean moons. But Amalthea likely did not form at its current orbital position as is suggested by the presence of hydrous minerals on its surface \citep{2004Sci...306.2224T}.} It has long been hypothesized that Jupiter's CPD dissipated when the ice line was between the orbits of Europa and Ganymede, at about 10 to $15\,R_{\rm Jup}$ \citep{1974Icar...21..248P}. Moreover, simulations of the orbital evolution of accreting proto-satellites in viscously dominated disks around Jupiter, Saturn, and Uranus indicate a universal scaling law for the total mass of satellite systems ($M_{\rm T}$) around the giant planets in the solar system \citep{2006Natur.441..834C,2010ApJ...714.1052S}, where $M_{\rm T}\approx10^{-4}$ times the planetary mass ($M_{\rm p}$).

\section{METHODS}
\label{sec:methods}

In the \citet{2002AJ....124.3404C} model, the accretion rate onto Jupiter was assumed to be time-independent. \citet{2006Natur.441..834C} focussed on the migration and growth of proto-moons, but they did not describe their assumptions for the temperature profile in the planetary accretion disk. Others used analytical descriptions for the temporal evolution of the accretion rates or for the movement of the H$_2$O ice lines \citep{1995SoSyR..29...85M,2004P&SS...52..361M,2006Natur.441..834C,2010ApJ...714.1052S,2012ApJ...753...60O} or they did not consider all the energy inputs described above \citep{2005A&A...439.1205A}.

We here construct, for the first time, a semi-analytical model for the CPDs of Jovian and super-Jovian planets that is linked to pre-computed planet evolution tracks and that contains four principal contributions to the disk heating: (i) viscous heating, (ii) accretion onto the CPD, (iii) planetary irradiation, and (iv) heating from the ambient circumstellar nebula. Compared to previous studies, this setup allows us to investigate many scenarios with comparatively low computational demands, and we naturally track the radial movement of the H$_2$O ice line over time. This approach is necessary, because we do not know any extrasolar moons that could be used to calibrate analytical descriptions for movement of the ice line around super-Jovian planets. We focus on the large population of Jovian and super-Jovian planets at around 1\,AU from Sun-like stars, several dozens of which had their masses determined through the radial velocity technique as of today.

\subsection{Disk Model}

The disk is assumed to be axially symmetric and in hydrostatic equilibrium. We adopt a standard viscous accretion disk model \citep{2002AJ....124.3404C,2006Natur.441..834C}, parameterized by a viscosity parameter $\alpha$ ($10^{-3}$ in our simulations) \citep{1973A&A....24..337S}, that is modified to include additional sources of disk heating \citep{2014SoSyR..48...62M}. We consider dusty gas disks around young ($\approx~10^6$\,yr old), accreting giant planets with final masses beyond that of Jupiter ($M_{\rm Jup}$). These planets accrete gas and dust from the circumstellar disk. Their accretion becomes increasingly efficient, culminating in the so-called runaway accretion phase when their masses become similar to that of Saturn \citep{2009Icar..199..338L,2013A&A...558A.113M}. Once they reach about a Jovian mass (depending on their distance to the star, amongst others), they eventually open up a gap in the circumstellar disk and their accretion rates drop rapidly. Hence, the formation of moons, which grow from the accumulation of solids in the CPD, effectively stops at this point or soon thereafter. A critical link between planet and moon formation is the combined effect of various energy sources (see the four heating terms described above) on the temperature distribution in the CPD and the radial position of the H$_2$O ice line.

In our disk model, the radial extent of the inner, optically thick part of the CPD, where moon formation is suspected to occur, is set by the disk's centrifugal radius ($r_{\rm cf}$). At that distance to the planet, centrifugal forces on an object with specific angular momentum $j$ are balanced by the planet's gravitational force. Using 3D hydrodynamical simulations, \citet{2008ApJ...685.1220M} calculated the circumplanetary distribution of the angular orbital momentum in the disk and demonstrated the formation of an optically thick disk within about $30\,R_{\rm Jup}$ around the planet. An analytical fit to their simulations yields \citep{2008ApJ...685.1220M}

\begin{equation}\label{eq:sam}
j(t) = \ 
\begin{cases}
\ 7.8 \times 10^{11} \left(  \frac{\displaystyle M_\mathrm{p}(t)}{\displaystyle M_\mathrm{Jup}} \right) \left(  \frac{\displaystyle a_\mathrm{{\star}p}}{\displaystyle 1\,\mathrm{AU}} \right)^{7/4} \mathrm{m}^2\,\mathrm{s}^{-1} \\
\hspace{4.1cm} \mathrm{for} \ M_\mathrm{p} < M_\mathrm{Jup} \\
\\
\ 9.0 \times 10^{11} \left(  \frac{\displaystyle M_\mathrm{p}(t)}{\displaystyle M_\mathrm{Jup}} \right)^{2/3} \left(  \frac{\displaystyle a_\mathrm{{\star}p}}{\displaystyle 1\,\mathrm{AU}} \right)^{7/4} \mathrm{m}^2\,\mathrm{s}^{-1} \\
\hspace{4.1cm} \mathrm{for} \ M_\mathrm{p} \geq M_\mathrm{Jup} \ \ ,
\end{cases}
\vspace{0.2cm}
\end{equation}

\noindent
where we introduced the variable $t$ to indicate that the planetary mass ($M_{\rm p}$) evolves in time. The centrifugal radius is then given by

\begin{equation}\label{eq:r_c}
r_{\rm cf} = \frac{j^2}{GM_\mathrm{p}} \ \ ,
\end{equation}

\noindent
with $G$ as Newton's gravitational constant. For Jupiter, this yields a centrifugal radius of about $22\,R_{\rm Jup}$, which is slightly less than the orbital radius of the outermost Galilean satellite, Callisto, at roughly $27\,R_{\rm Jup}$. Part of this discrepancy is due to thermal effects that are neglected in the \citep{2008ApJ...685.1220M} disk model. \citet{2009MNRAS.392..514M} investigated these thermal effects on the centrifugal disk size by comparing isothermal with adiabatic disk models. They found that adiabatic models typically yield larger specific angular momentum at a given planetary distance, which then translate into larger centrifugal disk radii that nicely match the width of Callisto's orbit around Jupiter.\footnote{In particular, their isothermal model M1I, used to fit our Eq.~(\ref{eq:sam}), has a radial specific momentum distribution that is about 1.1 times smaller at Callisto's orbital radius than their adiabatic model M1A2. This offset means an $(1.1)^2~=~1.21$-fold increase of the centrifugal radius, which nicely fits to our correction factor of $27/22~\approx~1.23$.} We thus introduce a thermal correction factor of 27/22 to the right-hand side of Equation~(\ref{eq:r_c}) following \citet{2009MNRAS.392..514M}, and therefore include Callisto at the outer edge of the optically thick part of our disk model. We note, though, that this slight rescaling hardly affects the general results of our simulations.

Recent 3D global magnetohydrodynamical (MHD) simulations by \citet[][see their Sect.~6.3]{2013ApJ...779...59G} produce circumplanetary surface gas densities that agree much better with the ``gas-starved'' model of \citet{2006Natur.441..834C}, which our model is derived from, than with the ``minimum mass'' model of \citet{2003Icar..163..198M}.\footnote{More advanced 3D MHD simulations would need to take into account the actual formation of the planet \citep[assumed to be a sink particle by][]{2013ApJ...779...59G} and would require resolving the inner parts of the compact disk to test whether this argument holds in favor of the gas-starved model.} The latter authors argue that Callisto formed in the low-density regions of an extended CPD with high specific angular momentum after Jupiter opened up a gap in the circumstellar disk. In their picture, the young Callisto accreted material from orbital radii as wide as $150\,R_{\rm Jup}$. In our model, however, Callisto forms in the dense, optically thick disk, where we suspect most of the solid material to pile up. Simulations by \citet{2006Natur.441..834C} and \citet{2010ApJ...714.1052S} show that our assumption can well reproduce the masses and orbits of the Galilean moons.

The disk is assumed to be mostly gaseous with an initial dust-to-mass fraction $X$ \citep[set to 0.006 in our simulations,][]{2013ApJ...778...78H}. Although we do not simulate moon formation in detail, we assume that the dust would gradually build planetesimals, either through streaming instabilities in the turbulent disk \citep{2014prpl.conf..547J} or through accumulation within vortices \citep{2003ApJ...582..869K}, to name just two possible formation mechanisms. The disk is parameterized with a fixed Planck opacity ($\kappa_{\rm P}$) in any of our simulations, but we tested various values. The fraction of the planetary light that contributes to the heating of the disk surface is parameterized by a coefficient $k_{\rm s}$, typically between 0.1 and 0.5  \citep{2014SoSyR..48...62M}. This quantity must not be confused with the disk albedo, which can take values between almost 0 and 0.9, depending on the wavelength and the grain properties \citep{2001ApJ...553..321D}. The sound velocity in the hydrogen (H) and helium (He) disk gas usually depends on the mean molecular weight ($\mu$) and the temperature of the gas, but in the disk midplane it can be approximated \citep{2014MNRAS.440...89K} as $c_{\rm s}~=~1.9\,{\rm km\,s}^{-1}~\sqrt{T_{\rm m}(r)/1000\,{\rm K}}$ for midplane temperatures $T_{\rm m}\lesssim1000\,{\rm K}$. At these temperatures, ionization can be neglected and $\mu=2.34\,{\rm kg\,mol}^{-1}$. Further, the disk viscosity is given by $\nu~=~{\alpha}c_{\rm s}^2/\Omega_{\rm K}(r)$, with $\Omega_{\rm K}(r)~=~\sqrt{GM_{\rm p}/r^3}$ as the Keplerian orbital frequency.

The steady-state gas surface density ($\Sigma_{\rm g}$) in the optically thick part of the disk can be obtained by solving the continuity equation for the infalling gas at the disk's centrifugal radius \citep{2006Natur.441..834C}, which yields

\begin{equation}\label{eq:Sigma}
\Sigma_\mathrm{g}(r) =  \frac{\dot{M}}{3\pi\nu} \, \times \, \frac{\Lambda(r)}{l}
\end{equation}

\noindent
where

\begin{align}\label{eq:Lambda} \nonumber
\Lambda(r) &= 1 - \frac{4}{5} \, \sqrt{\frac{r_\mathrm{cf}}{r_\mathrm{d}}} - \frac{1}{5} \, \left( \frac{r}{r_c} \right)^2 \\ \nonumber \\
l &= 1 - \sqrt{\frac{R_\mathrm{p}}{r_\mathrm{d}}}
\end{align}

\noindent
is derived from a continuity equation for the infalling material and based on the angular momentum delivered to the disk \citep{2006Natur.441..834C,2014SoSyR..48...62M}, and $\dot{M}$ is the mass accretion rate through the CPD, assumed to be equal to the mass dictated by the pre-computed planet evolution models \citep{2013A&A...558A.113M}. We set $r_{\rm d}=R_{\rm H}/5$, which yields $r_{\rm d}\approx154\,R_{\rm Jup}$ for Jupiter \citep{2010ApJ...714.1052S,2014SoSyR..48...62M}.

The effective half-thickness of the homogeneous flared disk, or its scale height, is derived from the solution of the vertical hydrostatic balance equation as

\begin{equation}\label{eq:h}
h(r) = \frac{c_\mathrm{s}(T_\mathrm{m}(r)) r^{3/2}}{\sqrt{GM_\mathrm{p}}} \ \ .
\end{equation}

\noindent
We adopt the standard assumption of vertical hydrostatic balance in the disk and assume that the gas density ($\rho_{\rm g}$) in the disk decreases exponentially with distance from the midplane as per

\begin{equation}\label{eq:rhos}
\rho_\mathrm{g}(r) =\rho_0 \, \mathrm{e}^{\frac{-z_s^2}{2h(r)^2}}
\end{equation}

\noindent
where $z$ is the vertical coordinate and $\rho_0$ the gas density in the disk midplane. The gas surface density is given by vertical integration over $\rho(r,z)$, that is,

\begin{equation}\label{eq:Sigmag}
\Sigma_{\rm g}(r) = \int_{\rm -inf}^{\rm +inf} dz \ \rho(r,z) \ \ .
\end{equation}

\noindent
Inserting Equation~(\ref{eq:rhos}) into Equation~(\ref{eq:Sigmag}), the latter can be solved for $\rho_0$ and we obtain

\begin{equation}\label{eq:rho0}
\rho_0(r) = \sqrt{\frac{2}{\pi}} \frac{\Sigma_\mathrm{g}(r)}{2h(r)} \ \ ,
\end{equation}

\noindent
which only depends on the distance $r$ to the planet. At the radiative surface level of the disk, or photospheric height ($z_{\rm s}$), the gas density equals 

\begin{equation}\label{eq:rhosr}
\rho_\mathrm{s}(r) =\rho_0 \, \mathrm{e}^{\frac{-z_s^2}{2h(r)^2}}
\end{equation}

\noindent
where we calculate $z_{\rm s}$ as

\begin{equation}\label{eq:zs}
z_\mathrm{s}(r) = \mathrm{erf}^{-1}{\Big (}1-\frac{2}{3}\frac{2}{\Sigma_\mathrm{g}(r)\kappa_\mathrm{P}}{\Big )} \sqrt{2} h(r) \ \ .
\end{equation}

\noindent
The latter formula is derived using the definition of the disk's optical depth

\begin{equation}
\tau = \int_{z_\mathrm{s}}^{+\inf} \mathrm{d}z \ \kappa_\mathrm{P} \rho(r,z) = \kappa_\mathrm{P} \int_{z_\mathrm{s}}^{+\inf} \mathrm{d}z \  \rho(r,z)
\end{equation}

\noindent
and our knowledge of $\tau=2/3$ at the radiating surface level of the disk, which gives

\begin{align}\label{eq:tau} \nonumber
\frac{2}{3} =& \ \kappa_\mathrm{P} \int_{z_\mathrm{s}}^{+\inf} \mathrm{d}z \  \rho_{\rm g}(r,z) \\
\Leftrightarrow z_\mathrm{s}(r) =& \ \mathrm{erf}^{-1}{\Big (}1-\frac{2}{3}\sqrt{\frac{2}{\pi}}\frac{1}{h(r)\rho_0(r)\kappa_\mathrm{P}}{\Big )} \sqrt{2} h(r) \ \ .
\end{align}

\noindent
Using Equation~({\ref{eq:rho0}}) for $\rho_0(r)$ in Equation~(\ref{eq:tau}), we obtain Equation~(\ref{eq:zs}).

Following the semi-analytical disk model of \citet{2014SoSyR..48...62M}, the disk surface temperature is given by the energy inputs of various processes as per

\begin{align}\label{eq:T_s}\nonumber
T_\mathrm{s}(r) = {\Bigg (}  & \frac{1 + (2\kappa_\mathrm{P}\Sigma_\mathrm{g}(r))^{-1}}{\sigma_\mathrm{SB}} {\Big (} F_\mathrm{vis}(r) +  F_\mathrm{acc}(r) \\
                          &  +  k_\mathrm{s}F_\mathrm{p}(r) {\Big )} + T_\mathrm{neb}^4  {\Bigg )}^{1/4} \ \ ,
\end{align}

\noindent
where

\begin{align}\label{eq:Fa}\nonumber
F_\mathrm{vis}(r) &= \frac{3}{8\pi} \frac{\Lambda(r)}{l}\dot{M}\Omega_\mathrm{K}(r)^2 \\ \nonumber
F_\mathrm{acc}(r) &= \frac{X_\mathrm{d}\chi G M_\mathrm{p}\dot{M}}{4{\pi}r_\mathrm{cf}^2r} \ \mathrm{e}^{-(r/r_\mathrm{cf})^2} \\ \nonumber
F_\mathrm{p}(r) &=  L_\mathrm{p} \frac{\sin{\Big (}\zeta(r)r+\eta(r){\Big )}}{8{\pi}(r^2+z_\mathrm{s}^2)} \\
\end{align}

\noindent
are the energy fluxes from viscous heating, accretion onto the disk, and the planetary illumination, and $T_{\rm neb}$ denotes the background temperature of the circumstellar nebula (100\,K in our simulations). The geometry of the flaring disk is expressed by the angles

\begin{align}\label{eq:angles} \nonumber
\zeta(r) &= \arctan\left( \frac{4}{3\pi} \frac{R_\mathrm{p}}{\sqrt{r^2+z_\mathrm{s}^2}} \right) \\
\eta(r) & = \arctan\left( \frac{\mathrm{d}z_\mathrm{s}}{\mathrm{d}r} \right) - \arctan\left( \frac{z_\mathrm{s}}{r} \right) \ \ .
\end{align}

\noindent
Taking into account the radiative transfer within the optically thick disk with Planck opacity $\kappa_{\rm P}$, the midplane temperature can be estimated as \citep{2014SoSyR..48...62M}

\begin{align}\label{eq:T_m}\nonumber
T_\mathrm{m}(r)^5 - T_\mathrm{s}(r)^4 \, T_\mathrm{m}(r) = & \ \frac{12\mu\chi\kappa_\mathrm{P}}{2^9\pi^2\sigma_\mathrm{SB}R_\mathrm{g}\gamma} \times \frac{\dot{M}^2}{\alpha}\Omega_\mathrm{K}(r)^3 \\
 & \times \left( \frac{\Lambda(r)}{l} \right)^2 q_\mathrm{s}(r)^2 \ \ ,
\end{align}

\noindent
where $\sigma_{\rm SB}$ is the Stefan-Boltzmann constant, $R_{\rm Gas}$ the ideal gas constant, $\gamma=1.45$ the adiabatic exponent (or ratio of the heat capacities), and

\begin{equation}\label{eq:q_s}\nonumber
q_\mathrm{s}(r) = 1 - \frac{4}{3\kappa_\mathrm{P}\Sigma_{\rm g}(r)}
\end{equation}

\noindent
is the vertical mass coordinate at $z_{\rm s}$.

\subsection{Planet Evolution Tracks}

%**********************************************
%Fig. 1
\begin{figure}[t]
  \centering
  \vspace{.0cm}
  \scalebox{0.35}{\includegraphics{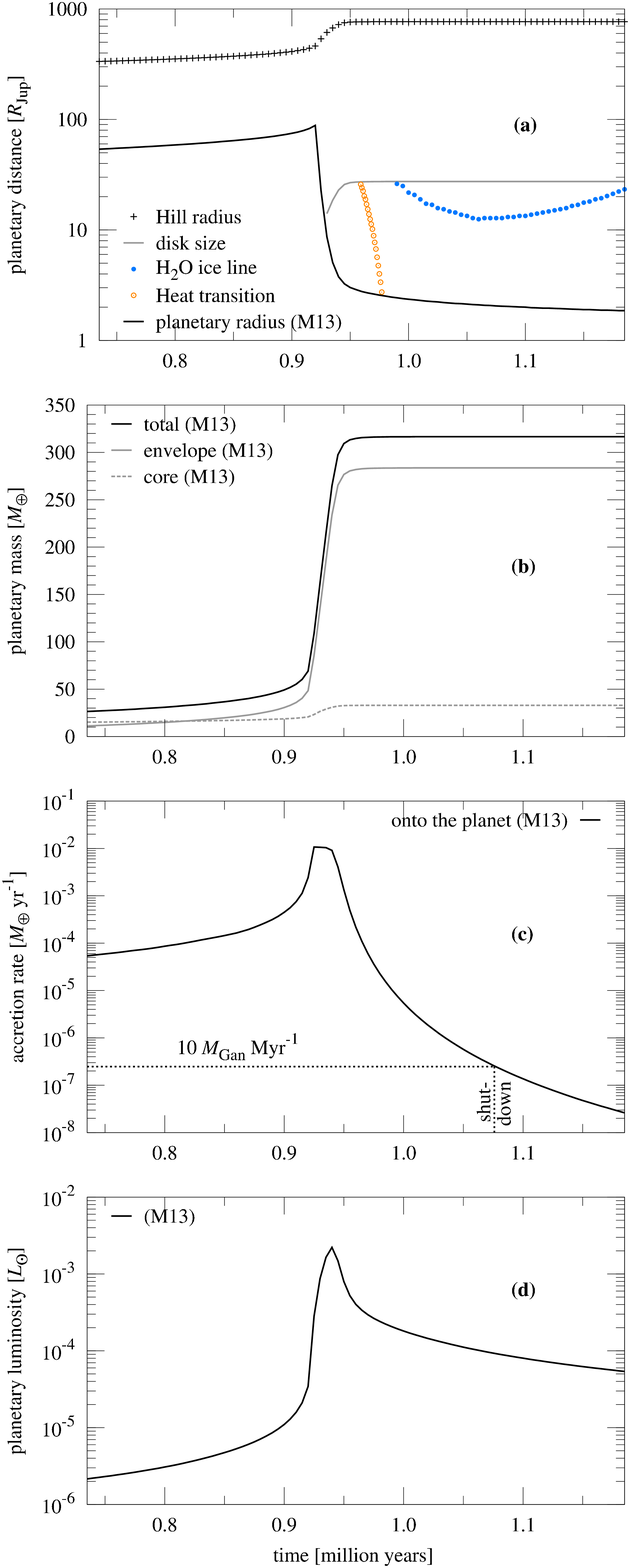}}
  \caption{Evolution of a Jupiter-like model planet and its circumplanetary disk. Values taken from \citet{2013A&A...558A.113M} are labeled ``M13''. \textbf{(a)} Circumplanetary disk properties. \textbf{(b)} Growth of the solid core, gaseous envelope, and total mass. \textbf{(c)} Total mass accretion rate. The dashed horizontal line indicates our fiducial shutdown rate for moon formation of $10\,M_{\rm Gan}\,{\rm Myr}^{-1}$. The dashed vertical line marks the corresponding shutdown for moon formation at about $1.08\times10^6$\,yr. \textbf{(d)} Planetary luminosity evolution.}
%     \vspace{.15cm}
  \label{fig:tracks}
\end{figure}
%**********************************************

We use a pre-computed set of planet formation models by \citet{2013A&A...558A.113M} to feed our planet disk model with the fundamental planetary properties such as the planet's evolving radius ($R_{\rm p}$), its mass, mass accretion rate ($\dot{M}_{\rm p}$), and luminosity ($L_{\rm p}$). Figure~\ref{fig:tracks} shows the evolution of these quantities with black solid lines indicating an accreting gas giant that ends up with one Jupiter mass or about 318 Earth masses ($M_\oplus$). In total, we have seven models at our disposal, where the planets have final masses of 1, 2, 3, 5, 7, 10, and $12\,M_{\rm Jup}$. These tracks are sensitive to the planet's core mass, which we assume to be $33\,M_\oplus$ for all planets. Jupiter's core mass is actually much lower, probably around $10\,M_\oplus$ \citep{1997Icar..130..534G}. Lower final core masses in these models translate into lower planetary luminosities at any given accretion rate. In other words, our results for the H$_2$O ice lines around the Jupiter-mass test planet are actually upper, or outer limits, and a more realistic evaluation of the conditions around Jupiter would shift the ice lines closer to the planet. Over the whole range of available planet tracks with core masses between 22 and $130\,M_\oplus$, we note that the planetary luminosities at shutdown are $10^{-4.11}$ and $10^{-3.82}$ solar luminosities, respectively. As the distance of the H$_2$O ice line in a radiation-dominated disk scales with $L_{\rm p}^{1/2}$, different planetary core masses would thus affect our results by less than ten percent.

The pre-computed planetary models cover the first few $10^6$\,yr after the onset of accretion onto the planet. We interpolate all quantities on a discrete time line with a step size of 5,000\,yr. At any given time, we feed Equations~(\ref{eq:Fa}) with the planetary model and solve the coupled Equations~(\ref{eq:sam})-(\ref{eq:zs}) in an iterative framework. With $T_{\rm s}$ provided by Equation~(\ref{eq:T_s}), we finally solve the 5th order polynomial in Equation~(\ref{eq:T_m}) numerically.

Once the planetary evolution models indicate that $\dot{M}_{\rm p}$ has dropped below a critical shutdown accretion rate ($\dot{M}_{\rm shut}$), we assume that the formation of satellites has effectively stopped. As an example, no Ganymede-sized moon can form once $\dot{M}_{\rm shut}~<~M_{\rm Gan}\,{\rm Myr}^{-1}$ ($M_{\rm Gan}$ being the mass of Ganymede) and if the disk's remaining life time is $<10^6$\,yr (see Figure~\ref{fig:tracks}c). As $\dot{M}_{\rm p}$ determines the gas surface density through Equation~(\ref{eq:Sigma}), different values for $\dot{M}_{\rm shut}$ mean different distributions of $\Sigma_\mathrm{g}(r)$. In particular, $\Sigma_{\rm g}(r~=~10\,R_{\rm Jup})$ equals $7.4~\times~10^2$, $9.7~\times~10^1$, and $7.8~\times~10^0\,{\rm kg\,m}^{-2}$ once $\dot{M}_{\rm shut}$ reaches 100, 10, and 1 $M_{\rm Gan}\,{\rm Myr}^{-1}$, respectively, for the planet that ends up with one Jupiter mass \citep[see Figure~4 in][]{2015arXiv150401668H}.

In any single simulation run, $\kappa_{\rm P}$ is assumed to be constant throughout the disk, and simulations of the planetary H$_2$O ice lines are terminated once the planet accretes less than a given $\dot{M}_{\rm shut}$. To obtain a realistic picture of a broad range of hypothetical exoplanetary disk properties, we ran a suite of randomized simulations, where $\kappa_{\rm P}$ and $\dot{M}_{\rm shut}$ were drawn from a lognormal probability density distribution. For $\log_{10}(\kappa_{\rm P}/[{\rm m}^2\,{\rm kg}^{-1}])$ we assumed a mean value of $-2$ and a standard variation of 1, and concerning the shutdown accretion rate we assumed a mean value of 1 for $\log_{10}(\dot{M}_{\rm shut}/[M_{\rm Gan}\,{\rm Myr}^{-1}])$ and a standard variation of 1.

To get a handle on the plausible surface absorptivities of various disk, we tested different values of $k_{\rm s}$ between 0.1 and 0.5. For each of the seven test planets, we performed 120 randomized simulations of the disk evolution and then calculated the arithmetic mean distance of the final water ice line. The resulting distributions are skewed and non-Gaussian. Hence, we compute both the downside and the upside semi standard deviation (corresponding to $\sigma/2$), that is, the distance ranges that comprise $+68.3\,\%/2=+34.15\,\%$ and $-34.15\,\%$ of the simulations around the mean. We also calculate the $2\sigma$ semi-deviation, corresponding to $+95.5\,\%/2=+47.75\,\%$ and $-47.75\,\%$ around the mean. Downside and upside semi deviations combined deliver an impression of the asymmetric deviations from the mean, and their sum equals that of the Gaussian standard deviations.

The total, instantaneous mass of solids in the disk at the time of moon formation shutdown is given as

\begin{equation}
M_{\rm s} = 2{\pi}X{\Bigg (}\int_{r_\mathrm{in}}^{r_\mathrm{ice}} \mathrm{d}r \ r \Sigma_\mathrm{g}(r) + 3\int_{r_\mathrm{ice}}^{r_\mathrm{cf}} \mathrm{d}r \ r \Sigma_\mathrm{g}(r) {\Bigg )} \ \ ,
\end{equation}

\noindent
where $r_{\rm in}$ is the inner truncation radius of the disk \citep[assumed at Jupiter's corotation radius of 2.25 $R_{\rm Jup}$,][]{2002AJ....124.3404C}, $r_{\rm ice}$ is the distance of the H$_2$O ice line, and $r_{\rm cf}$ is the outer, centrifugal radius of the disk \citep{2008ApJ...685.1220M}. Depending on the density of the disk gas and the size of the solid grains, the water sublimation temperature can vary by several degrees Kelvin \citep{1972Icar...16..241L,2006ApJ...640.1115L}, but we adopt 170\,K as our fiducial value \citep{2013ApJ...778...78H}.

We simulate the evolution of the H$_2$O ice lines in the disks around young super-Jovian planets at 5.2 astronomical units (AU, the distance between the Sun and the Earth) from a Sun-like star, facilitating comparison of our results to the Jovian moon system. These planets belong to the observed population of super-Jovian planets at $\approx1$\,AU around Sun-like stars \citep{2011MNRAS.417.1236H,2012ApJ...760..117H,2013Sci...340..572H}, and it has been shown that their satellite systems may remain intact during planet migration \citep{2010ApJ...719L.145N}.

%**********************************************
%Fig. 2
\begin{figure}[t]
  \centering
  \vspace{.0cm}
  \scalebox{0.56}{\includegraphics{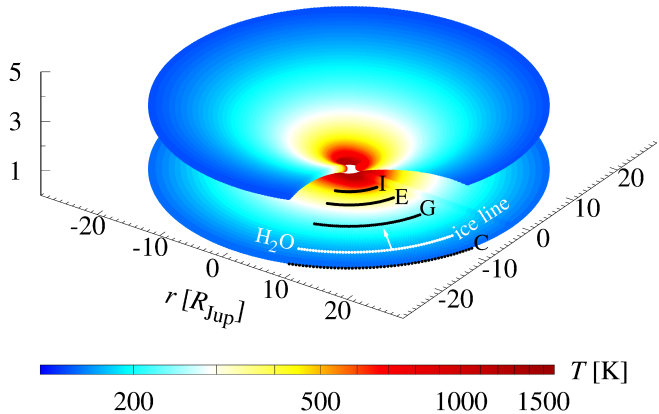}}
  \caption{Disk temperatures around a forming Jupiter-like planet $10^6$\,yr after the onset of accretion. The two disk levels represent the midplane and the photosphere (at a hight $z_{\rm s}$ above the midplane). At a given radial distance ($r$) to the planet, measured in Jupiter radii, the midplane is usually warmer than the surface (see color bar). The orbits of Io, Europa, Ganymede, and Callisto (labelled I, E, G, and C, respectively) and the location of the instantaneous H$_2$O ice line at $\approx22.5\,R_{\rm Jup}$ are indicated in the disk midplane. The arrow attached to the ice line indicates that it is still moving inward before it reaches its final location at roughly the orbit of Ganymede. This simulation assumes fiducial disk values ($k_{\rm s}=0.2, \kappa_{\rm P}=10^{-2}\,{\rm m}^2\,{\rm kg}^{-1}$), some $10^5$\,yr before the shutdown of moon formation.}
     \vspace{.15cm}
  \label{fig:disk}
\end{figure}
%**********************************************

\section{Results and Predictions}
\label{sec:results} 

Figure~\ref{fig:tracks}(a) shows, on the largest radial scales, the Hill radius (black crosses) of the accreting giant planet. The planetary radius (black solid line) is well within the Hill sphere, but it is quite extensive for  0.9\,Myr, so much that the CPD (gray solid line) has not yet formed by that time. It only appears after 0.9\,Myr of evolution of the system. Within that disk, we follow the time evolution of two features -- the heat transition (orange open circles) and the H$_2$O ice line (blue dots). The heat transition denotes the transition from the viscous to the irradiation heating regime in the disk \citep[see][]{2011MNRAS.417.1236H}, and it appearsÊat the outer disk edge about $0.95\times10^6$\,yr after the onset of accretion. It moves rapidly inwards and within $\approx2\times10^4$\,yr it reaches the inner disk edge, which sits roughly at the radius of the planet. At the same time ($\approx0.99\times10^6$\,yr after the onset of accretion), the H$_2$O ice line appears at the outer disk radius and then moves slowly inward as the planet cools. The ice line reverses its direction of movement at $\approx1.1\times10^6$\,yr due to the decreasing gas surface densities, while the opacities are assumed to be constant throughout the disk, see Equation~(\ref{eq:T_s}).

Figure~\ref{fig:tracks}(b) displays the mass evolution of the planetary core (gray dashed line) and atmosphere (gray solid line). Note that the rapid accumulation of the envelope and the total mass at around $0.93\times10^6$\,yr corresponds to the runaway accretion phase. Panel (c) presents the total mass accretion rate onto the planet (black solid line). The dashed horizontal line shows an example for $\dot{M}_{\rm shut}$ (here $10\,M_{\rm Gan}\,{\rm Myr}^{-1}$), which corresponds to a time 1.08\,Myr after the onset of accretion in that particular model. Note that shutdown accretion rates within one order of magnitude around this fiducial value occur 0.1 - 0.3\,Myr after the runaway accretion phase, that is, after the planet has opened up a gap in the circumstellar disk. In panel (d), the planetary luminosity peaks during the runaway accretion phase and then dies off as the planet opens up a gap in the circumstellar disk, which starves the CPD.

%**********************************************
%Fig. 3
\begin{figure}[t]
  \centering
  \scalebox{0.6}{\includegraphics{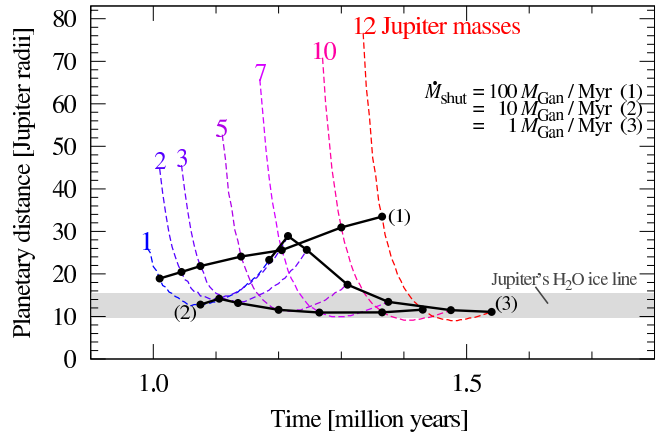}}
  \caption{Evolution of the H$_2$O ice lines in the disks around super-Jovian gas planets. Black solid lines, labeled (1) - (3), indicate the locations of the H$_2$O ice lines assuming different mass accretion rates for the shutdown of moon formation ($\dot{M}_{\rm shut}\in\{100,10,1\}{\times}M_{\rm Gan}\,{\rm Myr}^{-1}$). The shaded area embraces the orbits of Europa and Ganymede around Jupiter, where Jupiter's H$_2$O ice line must have been at the time when the Galilean satellites completed formation. The most plausible shutdown rate for the Jovian system (black line with label 2) predicts ice lines between roughly 10 and 15\,$R_{\rm Jup}$ over the whole range of super-Jovian planetary masses. Simulations assume $k_{\rm s}=0.2$ and $\kappa_{\rm P}=10^{-2}\,{\rm m}^2\,{\rm kg}^{-2}$.}
  \label{fig:icelines}
\end{figure}
%**********************************************

Figure~\ref{fig:disk} shows a snapshot of the temperature structure of the disk surface and midplane around a Jupiter-mass planet, $10^6$\,yr after the onset of mass accretion. The location of the instantaneous water ice line is indicated with a white dotted line, and the current positions of the Galilean satellites are shown with black dotted lines. Over the next hundred thousand years, the heating rates drop and the ice line moves inward to Ganymede's present orbit as the planet's accretion rate decreases due to its opening of a gap in the circumsolar disk. We argue that the growing Ganymede moved with the ice line trap and was parked in its present orbit when the circumjovian disk dissipated. Due to the rapid decrease of mass accretion onto the planet after gap opening (up to about an order of magnitude per $10^4$\,yr), this process can be reasonably approximated as an instant shutdown on the time scales of planet formation (several $10^6$\,yr), although it is truly a gradual process.

Figure~\ref{fig:icelines} shows the radial positions of the ice lines around super-Jovian planets as a function of time and for a given disk surface absorptivity ($k_{\rm s}$) and disk Planck opacity ($\kappa_{\rm P}$). More massive planets have larger disks and are also hotter at a given time after the onset of accretion, which explains the larger distance and later occurrence of water ice around the more massive giants. Solid black lines connect epochs of equal accretion rates (1, 10, and $100\,M_{\rm Gan}$ per Myr). Along any given ice line track, higher accretion rates correspond to earlier phases. The gray shaded region embraces the orbital radii of Europa and Ganymede, between which we expect the H$_2$O ice line to settle. The H$_2$O ice line around the 1 Jupiter mass model occurs after $\approx0.99\times10^6$\,yr at the outer edge of the disk, passes through the current orbit of Ganymede, and then begins to move outwards around $1.1\times10^6$\,yr due to the decreasing gas surface densities (note, the opacities are assumed constant). In this graph, $\dot{M}_{\rm shut}\approx10\,M_{\rm Gan}\,{\rm Myr}^{-1}$ can well explain the mentioned properties in the Galilean system.

\begin{figure}[t]
  \centering
  \scalebox{0.59}{\includegraphics{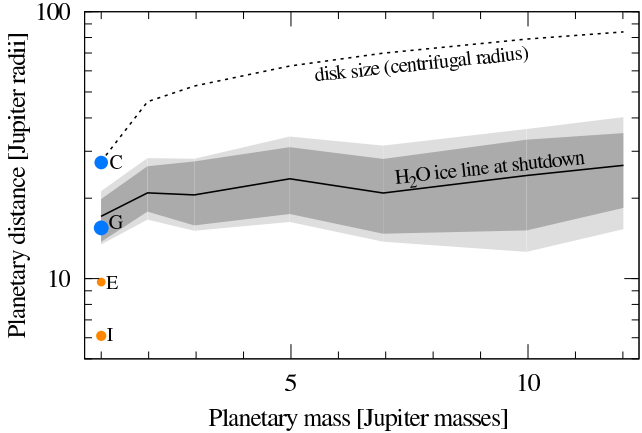}}
  \caption{Distance of the H$_2$O ice lines at the shutdown of moon formation around super-Jovian planets. The solid line indicates the mean, while shaded areas denote the statistical scatter (dark gray $1\,\sigma$, light gray $2\,\sigma$) in our simulations, based on the posterior distribution of the disk Planck mean opacity ($\kappa_{\rm P}$) and the shutdown accretion rate for moon formation ($\dot{M}_{\rm shut}$). The dashed line represents the size of the optically thick part of the circumplanetary disk, or its centrifugal radius. All planets are assumed to orbit a Sun-like star at a distance of 5.2\,AU and $k_{\rm s}$ is set to 0.2. Labeled circles at $1\,M_{\rm Jup}$ denote the orbits of the Galilean satellite Io, Europa, Ganymede, and Callisto. Orange indicates rocky composition, blue represents H$_2$O-rich composition. Circle sizes scale with moon radii. Note that Ganymede sits almost exactly on the circumjovian ice line.}
  \label{fig:icelines_mass}
\end{figure}

In Figure~\ref{fig:icelines_mass}, we present the locations of the ice lines in a more global picture, obtained by performing 120 randomized disk simulations for each planet, where $\dot{M}_{\rm shut}$ and $\kappa_{\rm P}$ were drawn from a lognormal probability distribution. We also simulated several plausible surface absorptivities of the disk \citep{2001ApJ...553..321D,2014SoSyR..48...62M} ($0.1{\leq}k_{\rm s}{\leq}0.5$), which resulted in ice line locations similar to those shown in Figure~\ref{fig:icelines_mass}, where $k_{\rm s}=0.2$. The mean orbital radius of the ice line at the time of shutdown around the $1\,M_{\rm Jup}$ planet is almost precisely at Ganymede's orbit around Jupiter, which we claim is no mere coincidence. Most importantly, despite a variation of $\dot{M}_{\rm shut}$ by two orders of magnitude and considering more than one order of magnitude in planetary masses, the final distances of the H$_2$O ice lines only vary between about 15 and $30\,R_{\rm Jup}$. Hence, regardless of the actual value of $\dot{M}_{\rm shut}$, the transition from rocky to icy moons around giant planets at several AU from Sun-like stars should occur  at planetary distances similar to the one observed in the Galilean system.

We ascribe this result to the fact that the planetary luminosity is the dominant heat source at the time of moon formation shutdown. Planetary luminosity, in turn, is determined by accretion (and gravitational shrinking), hence a given $\dot{M}_{\rm shut}$ translates into similar luminosities and similar ice line radii for all super-Jovian planets. Planets above $1\,M_{\rm Jup}$ have substantially larger parts of their disks beyond their water ice lines (note the logarithmic scale in Figure~\ref{fig:icelines_mass}) and thus have much more material available for the formation of giant, water-rich analogs of Ganymede and Callisto.

\begin{figure}[t]
  \centering
  \scalebox{0.59}{\includegraphics{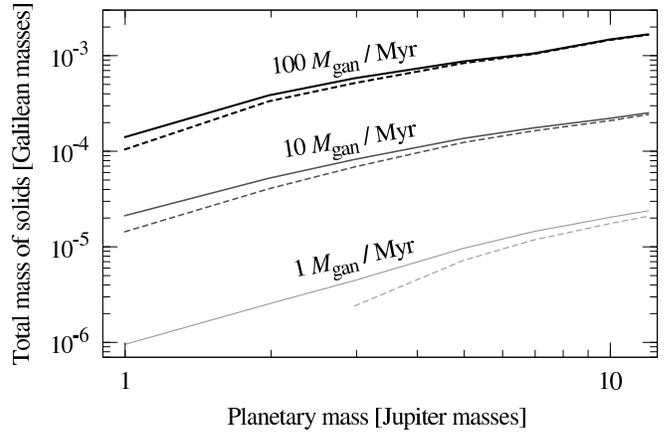}}
  \caption{Instantaneous mass of solids in the disks around super-Jovian planets at moon formation shutdown. Solid and dashed lines refer to disk absorptivities of $k_{\rm s}=0.2$ and 0.4, respectively. The ordinate scales in units of the total mass contained in the Galilean moons ($\approx2.65\,M_{\rm Gan}$). For any given shutdown rate, we find a linear increase in the mass of solids at the time of moon formation shutdown as a function of planetary mass, in agreement with previous simulations for the solar system giants \citep{2006Natur.441..834C,2010ApJ...714.1052S}. Super-Jovian planets of $10\,M_{\rm Jup}$ should thus have moon systems with total masses of $\approx10^{-3}{\times}M_{\rm Jup}$, or 3 times the mass of Mars.}
  \label{fig:total_mass}
\end{figure}

Figure~\ref{fig:total_mass} shows the total mass of solids at the time of moon formation shutdown around super-Jovian planets. Intriguingly, for any given shutdown accretion rate the total mass of solids scales proportionally to the planetary mass. This result is not trivial, as the mass of solids depends on the location of the H$_2$O ice line at shutdown. Assuming that $\dot{M}_{\rm shut}$ is similar among all super-Jovian planets, we confirm that the $M_{\rm T}\propto10^{-4}\,M_{\rm p}$ scaling law observed in the solar system also applies for extrasolar super-Jupiters \citep{2006Natur.441..834C,2010ApJ...714.1052S}.

In addition to the evolution of the H$_2$O ice lines, we also tracked the movements of the heat transitions, a specific location within the disk, where the heating from planetary irradiation is superseded by viscous heating. Heat transitions cross the disk within only about $10^4$\,yr (see Figure~\ref{fig:tracks}a), several $10^5$\,yr before the shutdown of moon formation, and thereby cannot possibly act as moon traps. Their rapid movement is owed to the abrupt starving of the planetary disk due to the gap opening of the circumstellar disk, whereas the much slower photoevaporation of the latter yields a much slower motion of the circumstellar heat trap. The ineffectiveness of heat traps for satellites reflects a key distinction between the processes of moon formation and terrestrial planet formation.

%**********************************************
%Fig. 4
%**********************************************

%**********************************************
%Fig. 5
%**********************************************

\section{Discussion}
\label{subsec:discussion}

\subsection{Accretion and Migration of both Planets and Moons}

While \citet{2002AJ....124.3404C} stated that accretion rates of $2\times10^{-7}\,M_{\rm Jup}\,{\rm yr}^{-1}$ (about $2.6\times10^{4}\,M_{\rm Gan}\,{\rm Myr}^{-1}$) best reproduced the disk conditions in which the Galilean satellites formed, our calculations predict a shutdown accretion rate that is considerably lower, closer to $10\,M_{\rm Gan}\,{\rm Myr}^{-1}$. The difference in these results is mainly owed to two facts. First, \citet{2002AJ....124.3404C} only considered viscous heating.\footnote{\citet{2002AJ....124.3404C} discuss the contribution of planetary luminosity to the disk's energy budget, but for their computations of the gas surface densities they ignore it.} Our additional heating terms (illumination from the planet, accretion onto the disk and stellar illumination) contribute additional heat, which imply smaller accretion rates to let the H$_2$O ice lines move close enough to the Jupiter-like planet. Second, the parameterization of planetary illumination in the \citet{2002AJ....124.3404C} model is different from ours. While they assume an $r^{-3/4}$ dependence of the midplane temperature from the planet ($r$ being the planetary radial distance), we do not apply any pre-described $r$-dependence. In particular, $T_{\rm m}(r)$ cannot be described properly by a simple polynomial due to the different slopes of the various heat sources as a function of planetary distance.

Previous models assume that type I migration of the forming moons leads to a continuous rapid loss of proto-satellites into the planet \citep{2002AJ....124.3404C,2003Icar..163..198M,2003Icar..163..232M,2005A&A...439.1205A,2010ApJ...714.1052S}. \citep{2005A&A...439.1205A} considered Jupiter's accretion disk as a closed system after the circumstellar accretion disk had been photo-evaporated, whereas \citet{2010ApJ...714.1052S} described accretion onto Jupiter with an analytical model. In the \citet{2003Icar..163..198M,2003Icar..163..232M}Ê(ME) model, satellites migrate via type I but perturb the gas as they migrate and eventually stall and open a gap, ensuring their survival. In opposition to the Canup \& Ward (CW) theory, their model does not postulate ``generations'' of satellites, which are subsequently lost into the planet, because satellite formation doesn't start until the accretion inflow onto the planet wanes.

There are two difficulties with the CW picture. First, type I migration can be drastically slowed down as growing giant moons get trapped by the ice lines or at the inner truncation radius of the disk\footnote{ An inner cavity can be caused by magnetic coupling between the rotating planet and the disk \citep{1996Icar..123..404T}, and it can be an important aspect to explain the formation of the Galilean satellites \citep{2010ApJ...714.1052S}}. Thus it is not obvious that a conveyor belt of moons into their host planets is ever established. Second, our Figure~\ref{fig:total_mass} also contradicts this scenario, because the instantaneous mass of solids in the disk during the end stages of moon formation (or planetary accretion) is not sufficient to form the last generation of moons. In other words, whenever the instantaneous mass of solids contained in the circumjovian disk was similar to the total mass of the Galilean moons, the correspondingly high accretion rates caused the H$_2$O ice line to be far beyond the orbits of Europa and Ganymede.

We infer, therefore, that the final moon population around Jupiter and other Jovian or super-Jovian exoplanets must, at least to a large extent, have built during the ongoing, final accretion process of the planet, when it was still fed from the circumstellar disk.\footnote{This conclusion is similar to that proposed by the ME model, but for reasons that are very different.} In order to counteract the inwards flow due to type I migration, we suggest a new picture in which the circumplanetary H$_2$O ice line and the inner cavity of Jupiter's accretion disk have acted as migration traps. This important hypothesis needs to be tested in future studies. The effect of an inner cavity will also need to be addressed, as it might have been essential to prevent Io and Europa from plunging into Jupiter.

In our picture, Io should have formed dry and its migration might have been stopped at the inner truncation radius of Jupiter's accretion disk, at a few Jupiter radii \citep{1996Icar..123..404T}. It did not form wet and then lose its water through tidal heating. Ganymede may have formed at the water ice line in the circumjovian disk, where it has forced Io and Europa in the 1:2:4 orbital mean motion resonance \citep{1829mecc.book.....L}. From a formation point of view, we suggest that Io and Europa be regarded as moon analogs of the terrestrial planets, whereas Ganymede and Callisto resemble the precursors of giant planets.

Our combination of planet formation tracks and a CPD model enables new constraints on planet formation from moon observations. As just one example, the ``Grand Tack'' (GT) model suggests that Jupiter migrated as close as about 1.5\,AU to the Sun before it reversed its migration due to a mutual orbital resonance with Saturn \citep{2011Natur.475..206W}. In the proximity of the Sun, however, solar illumination should have depleted the circumjovian accretion disk from water ices during the end stages of Jupiter's accretion \citep[][in prep.]{HMP2015}. {Thus, Ganymede and Callisto would have formed in a dry environment \textit{during} the GT}, which is at odds with their high H$_2$O ice contents. They can also hardly have formed over millions of years \citep{2003Icar..163..198M} thereafter, because Jupiter's CPD (now truncated from its environment by a gap) still would have been dry. Alternatively, one might suggest that Callisto and Ganymede formed \textit{after} the GT from newly accreted planetesimals into a still active, gaseous disk around Jupiter. But then Io and Europa might have been substantially enriched in water, too. \citet[][see their Figure~8]{2014ApJ...784..109T} found that planetesimal accretion via gas drag is most efficient between 0.005 and 0.001 Hill radii ($R_{\rm H}$) or about 4 to $8\,R_{\rm Jup}$ where gas densities are relatively high.

To come straight to the point, our preliminary studies suggest that in the GT paradigm, the icy Galilean satellites must have formed \textit{prior to} Jupiter's excursion to the inner solar system \citep[][in prep.]{HMP2015}. This illustrates the great potential of moons to constrain planet formation, which is particularly interesting for the GT scenario where the timing of migration and planetary accretion is yet hardly constrained otherwise \citep{2014arXiv1409.6340R}.

%The lifetimes of the circumplanetary disks may indeed be longer than the lifetimes of the circumstellar disks (Fujii et al. 2014).

\subsection{Parameterization of the Disk}

Finally, we must  address a technical issue, namely,  our choice of the $\alpha$ parameter ($10^{-3}$).  While this is consistent with many previous studies, how would a variation of $\alpha$ change our results? Magnetorotational instabilities might be restricted to the upper layers of CPDs, where they become sufficiently ionized (mostly by cosmic high-energy radiation and stellar X-rays). Magnetic turbulence and viscous heating in the disk midplane might thus be substantially lower than in our model \citep{2014ApJ...785..101F}. On the other hand, \citet{2013ApJ...779...59G} modeled the magnetic stresses in CPDs with a 3D magnetohydrodynamic model and inferred $\alpha$ values of 0.01 and larger, which would strongly enhance viscous heating. Obviously, sophisticated numerical simulations of giant planet accretion do not yet consistently describe the magnetic properties of the disks and the associated $\alpha$ values.

Given that circumstellar disks are almost certainly magnetized, CPD can be expected to have inherited magnetic fields from this source.  This makes it likely that magnetized disk winds can be driven off the CPD \citep{2003A&A...411..623F,2007prpl.conf..277P} which can carry significant amounts of angular momentum. Even in the limit of very low ionization, \citet{2013ApJ...769...76B} demonstrated that magnetized disk winds will transport disk angular momentum at the rates needed to allow accretion onto the central object. However, independent of these uncertainties, the final positions of the H$_2$O ice line produced in our simulations turn out to depend mostly on planetary illumination, because viscous heating becomes negligible almost immediately following gap opening. Hence, even substantial variations of $\alpha$ by a factor of ten would hardly change our results for the ice line locations at moon formation shutdown since these must develop in radiatively dominated disk structure (but it would alter them substantially in the viscous-dominated regime before and during runaway accretion).

Our assumption of a constant Planck opacity throughout the disk is simplistic and ignores the effects of grain growth, grain distribution within the disk, as well as the evolution of the disk properties. In a more consistent model, $\kappa_{\rm P}$ depends on both the planetary distance and distance from the midplane, which might entail significant modifications in the temperature distribution that we predict.

\section{Conclusions}
\label{subsec:conclusion}

We have demonstrated that ice lines imprint important structural features on systems of icy moons around massive planets. Given that observations show a strong concentration of super-Jovian planets at $\approx1$\,AU, we focused our analysis on the formation of massive moons in this planetary population.

After a forming giant planet opens up a gap in the circumstellar disk, its accretion rates and the associated viscous heating in the CPD drop substantially. We find that a heat transition crosses the CPD within $10^4$\,yr, which is too fast for it to act as a moon migration trap. Alternatively, we propose that moon migration can be stalled at the H$_2$O ice line, which moves radially on a $10^5$\,yr timescale. For Jupiter's final accretion phase, when the Galilean moons are supposed to form in the disk, our calculations show that the H$_2$O line is at about the contemporary radial distance of Ganymede, suggesting that the most massive moon in the solar system formed at a circumplanetary migration trap. Moreover, dead zones might be present in the inner CPD regions \citep{2013ApJ...779...59G} where they act as additional moon migration traps, but this treatment is beyond the scope of this paper.

Our model confirms the mass scaling law for the most massive planets, which suggests that satellite systems with total masses several times the mass of Mars await discovery. Their most massive members will be rich in water and possibly parked in orbits at their host planet's H$_2$O ice lines at the time of moon formation shutdown, that is, between 15 and $30\,R_{\rm Jup}$ from the planet. A Mars-mass moon composed of 50\,\% of water would have a radius of $\approx0.7$ Earth radii \citep{Fortney2007}. Although we considered giant planet accretion beyond 1\,AU, super-Jovian planets are most abundant around 1\,AU \citep{2013Sci...340..572H} and their moon systems have been shown to remain intact during planet migration \citep{2010ApJ...719L.145N}. Giant water-rich moons might therefore form an abundant population of extrasolar habitable worlds \citep{1997Natur.385..234W,2014arXiv1408.6164H} and their sizes could make them detectable around photometrically quiet dwarf stars with the transit method \citep{2012ApJ...750..115K,2014ApJ...787...14H}. In a few cases, the transits of such giant moons in front of hot, young giant planets might be detectable with the \textit{European Extremely Large Telescope}, with potential for follow-up observations of the planetary Rossiter-McLaughlin effect \citep{2014ApJ...796L...1H}.

More detailed predictions can be obtained by including the migration process of the accreting planet, which we will present in an upcoming paper. Ultimately, we expect that there will  be a competition between the formation of water-rich, initially icy moons beyond the circumplanetary H$_2$O ice line and the gradual heating of the disk (and loss of ices) during the planetary migration towards the star. Such simulations have the potential to generate a moon population synthesis with predictions for the abundance and detectability of large, water-rich moons around super-Jovian planets.

\acknowledgments

The report of an anonymous referee helped us to clarify several aspects of the manuscript. We thank C.~Mordasini for sharing with us his planet evolution tracks, G.~D'Angelo for discussions related to disk ionization, A.~Makalkin for advice on the disk properties, and Y.~Hasegawa for discussions of ice lines. R.~Heller is supported by the Origins Institute at McMaster University and by the Canadian Astrobiology Program, a Collaborative Research and Training Experience Program funded by the Natural Sciences and Engineering Research Council of Canada (NSERC). R.~E.~Pudritz is supported by a Discovery grant from NSERC.

%BIBLIOGRAPHY
\bibliography{ms}
\bibliographystyle{apj2}

\end{document}